\documentclass{article}
\PassOptionsToPackage{table}{xcolor}

\usepackage{graphicx}
\usepackage[T1]{fontenc}

\usepackage{graphicx,verbatim}
\usepackage{authblk}
\usepackage{hyperref}
\usepackage{booktabs}
\usepackage{tikz}
\usepackage{adjustbox}
\usepackage{pifont}
\usepackage{multirow}
\usepackage{tabularx}
\usepackage{caption}
\usepackage{subcaption}
\usepackage[table]{xcolor}

\newcommand*{\our}{\textbf{Histo-VL}}

\title{How Good is my Histopathology Vision-Language Foundation Model? A Holistic Benchmark}
\author[1]{Roba Al Majzoub}
\author[1]{Hashmat Malik}
\author[2]{Muzammal Naseer}
\author[1]{Zaigham Zaheer}
\author[3]{Tariq Mahmood}
\author[1]{Salman Khan}
\author[1]{Fahad Khan}

\affil[1]{Department of Computer Vision, Mohamed Bin Zayed University of Artificial Intelligence}
\affil[2]{Khalifa University}
\affil[3]{Shaukat Khanum Cancer Hospital}

\date{}

\begin{document}
\maketitle
\begin{abstract}
Recently, histopathology vision-language foundation models (VLMs) have gained popularity due to their enhanced performance and generalizability across different downstream tasks. However, most existing histopathology benchmarks are either unimodal or limited in terms of diversity of clinical tasks, organs, and acquisition instruments, as well as their partial availability to the public due to patient data privacy. As a consequence, there is a lack of comprehensive evaluation of existing histopathology VLMs on a unified benchmark setting that better reflects a wide range of clinical scenarios. To address this gap, we introduce \our{}, a fully open-source comprehensive benchmark comprising images acquired using up to 11 various acquisition tools that are paired with specifically crafted captions by incorporating class names and diverse pathology descriptions. Our \our{} includes 26 organs, 31 cancer types, and a wide variety of tissue obtained from 14 heterogeneous patient cohorts, totaling more than 5 million patches obtained from over 41K WSIs viewed under various magnification levels. We systematically evaluate existing histopathology VLMs on \our{} to simulate diverse tasks performed by experts in real-world clinical scenarios. Our analysis reveals interesting findings, including large sensitivity of most existing histopathology VLMs to textual changes with a drop in balanced accuracy of up to 25\% in tasks such as Metastasis detection, low robustness to adversarial attacks, as well as improper calibration of models evident through high ECE values and low model prediction confidence, all of which can affect their clinical implementation. Our benchmark and analysis code is available at our \href{https://github.com/musk007/Histopathology_Benchmark}{Github repository}.

\end{abstract}

\section{Introduction}
\label{section1}
Histopathological image assessment, a vital inspection that oncologists use to diagnose, stage, and plan cancer treatment, is a highly challenging field due to multiple reasons, including high dimensionality of images and significant observer disagreement. This motivated researchers to develop deep learning (DL) systems to help clinicians diagnose and predict cancer outcomes.
The recent popularity of vision-language models (e.g., CLIP~\cite{clip} and ALIGN~\cite{align}) trained on large cohorts of natural image-text pairs encouraged their application in healthcare. These vision-language foundation models (VLMs), were trained in an unsupervised manner on massive image-text pairs of data. Recent medical VLMs were trained on large-scale online-scraped data like BiomedCLIP~\cite{biomedclip}; similarly, the approach was adopted in histopathology to create VLMs like PLIP~\cite{plip} and QuiltNet~\cite{quilt}. To further refine the learned representations, a combination of online data and data from hospitals was used for better performance, as in CONCH~\cite{conch} and MI-Zero~\cite{miZero}, showing promising improvements on downstream tasks, such as cancer detection and survival prediction.

To evaluate histopathology DL models, several attempts in the literature have been made to test their efficacy, including cancer-dedicated challenges like PatchCamelyon~\cite{pcam}, PAIP~\cite{paip}, and BACH~\cite{bach}. Since these challenges do not capture the full diversity required for medical model assessment, recent works benchmarked histopathology vision foundation models on cancer detection~\cite{campanella2024clinical}, cancer subtyping~\cite{breen2024histopathology}, some even tested the models on a combination of tasks such as cancer subtyping and mutation prediction~\cite{neidlinger2024benchmarking}, or cancer subtyping and detection~\cite{kang2023benchmarking}. 

\begin{figure}[t!]
\centering
\includegraphics[width=\textwidth]{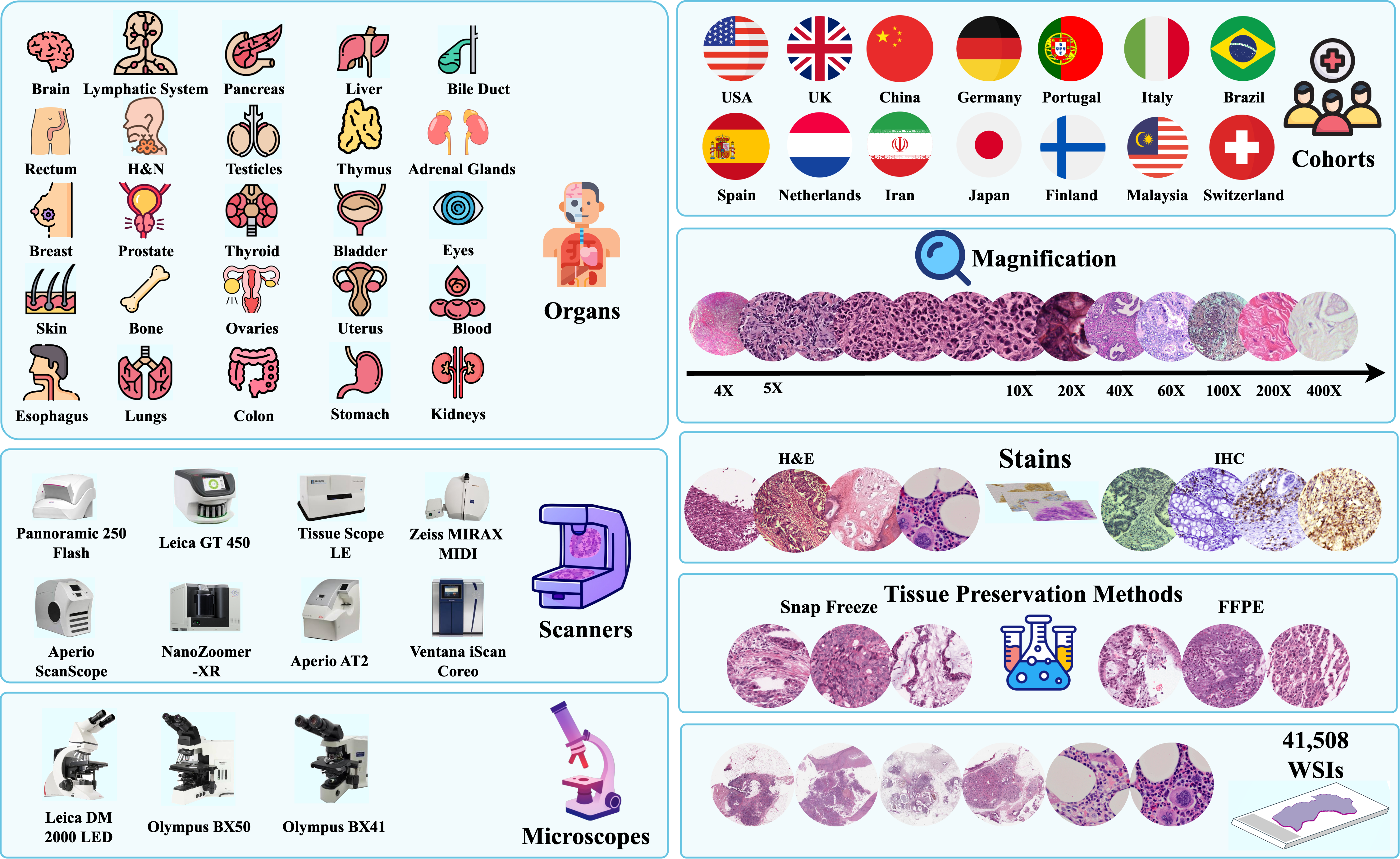}
\caption{\our{}, is a comprehensive benchmark encompassing cancers from 26 organs represented across 41,508 WSIs. These slides have undergone various tissue preservation methods and are acquired using 11 distinct imaging devices, including microscopes and scanners. Data is diverse, featuring cohorts from 14 countries worldwide. Best viewed zoomed in.}
\label{benchProperties}
\end{figure}

\begin{table}[t!]
\centering
\tiny
\resizebox{\textwidth}{!}{%
\begin{tabular}{l|c|c|c|c|c}
\hline
\textbf{Benchmark} &\textbf{Kang~\cite{kang2023benchmarking}} &\textbf{Campanella~\cite{campanella2024clinical}} & \textbf{Neidlinger~\cite{neidlinger2024benchmarking}} &  \textbf{Breen~\cite{breen2024histopathology}} & \textbf{Ours}  \\ \hline

Detection & \textcolor{green!60!black}{\ding{51}} & \textcolor{green!60!black}{\ding{51}} & \textcolor{red!70!black}{\ding{55}} & \textcolor{red!70!black}{\ding{55}} & \textcolor{green!60!black}{\ding{51}}\\ 
Subtyping & \textcolor{green!60!black}{\ding{51}} & \textcolor{red!70!black}{\ding{55}} & \textcolor{green!60!black}{\ding{51}}  & \textcolor{green!60!black}{\ding{51}}  &\textcolor{green!60!black}{\ding{51}}\\ 
Grading & \textcolor{red!70!black}{\ding{55}} & \textcolor{red!70!black}{\ding{55}} & \textcolor{red!70!black}{\ding{55}} & \textcolor{red!70!black}{\ding{55}}  & \textcolor{green!60!black}{\ding{51}}\\ 
Mutation prediction & \textcolor{red!70!black}{\ding{55}} & \textcolor{red!70!black}{\ding{55}} & \textcolor{green!60!black}{\ding{51}} & \textcolor{red!70!black}{\ding{55}} & \textcolor{green!60!black}{\ding{51}}\\ 

VLMs & \textcolor{red!70!black}{\ding{55}} & \textcolor{red!70!black}{\ding{55}} & \textcolor{red!70!black}{\ding{55}} & \textcolor{red!70!black}{\ding{55}} &\textcolor{green!60!black}{\ding{51}}\\ 
cancer types & 3 & 8 & 4 & 1 &  26\\ 
WSI & 860 & 19,837 & 9,493 & 693 &  41,508\\ 
Open-source & \textcolor{red!70!black}{\ding{55}} & \textcolor{green!60!black}{\ding{51}}(partial) & \textcolor{green!60!black}{\ding{51}}(partial) &  \textcolor{green!60!black}{\ding{51}} & \textcolor{green!60!black}{\ding{51}}\\ 
\bottomrule
\bottomrule
\end{tabular}
}
\caption{Comparison of \our{} with current histopathology benchmarks. \our{} covers more than five clinical tasks, including cancer detection, subtyping, and grading. The multimodal setting of image-text pairs, with patches derived from 41,508 WSIs, allows benchmarking of histopathology VLMs, on 26 main cancer types. Additionally, \our{} is fully open-sourced, since images are acquired from openly licensed research sources.}
\label{tab:benchmarks}
\end{table}

While these benchmarks typically evaluate histopathology VLMs on different cancer types through incorporating WSIs, they are limited in terms of clinical tasks, organs, and acquisition instruments, and most are partially available for the public due to patient data privacy (see Table~\ref{tab:benchmarks}).
Moreover, there is a lack of comprehensive evaluation of existing histopathology VLMs on a unified benchmark setting. Motivated by these observations in Table~\ref{tab:benchmarks}, we look into developing a comprehensive open-source benchmark that covers a wide range of clinical settings to extensively evaluate histopathology VLMs.
To this end we introduce a comprehensive benchmark, named \our{}  that includes 26 organs, represented by 32 cancer types, obtained from 14 heterogeneous patient cohorts, as well as images captured using up to 11 various acquisition tools, and corresponding texts that caption the pathology depicted in the images.
Our benchmark is meticulously curated to represent a large number of tissues from various cohorts, including Japan, Iran, Switzerland, and USA. Tissues are processed using multiple tissue preservation techniques like FFPE and snap freezing, and viewed under various magnification levels (4X-400X), resulting in patches amounting to more than 5 million in number, as shown in Figure~\ref{benchProperties} and Table~\ref{tab:dataCharacteristics}. 
This not only enables the evaluation of the performance of histopathology VLMs under different distribution shifts and in a wider range of clinical settings but also helps analyze the different challenges that a histopathology VLM faces in real-world applications.
In summary, our main contributions are as follows:
\begin{itemize}
    \item We introduce \our{} consisting of images paired with specifically crafted captions that incorporate class names and diverse descriptions to extensively evaluate histopathology VLMs. The proposed benchmark comprises 26 organs, 14 different cohorts, more than 5 million patches obtained from over 41K WSIs, and 11 different acquisition tools (see Figure~\ref{benchProperties}).
    \item We extensively evaluate existing histopathology VLMs through a series of experiments targeting key clinical tasks performed by experts, ensuring a better reflection of models' performances in real-world clinical scenarios. Our analysis reveals several interesting findings, including the sensitivity of existing models to textual changes with a drop in balanced accuracy of up to 25\% in tasks such as Metastasis detection. Moreover, we observe a lack of proper calibration, which is evident in the high ECE values and low confidence that are likely to affect adoption of models in clinical settings.
\end{itemize}

\section{Proposed \our{} Benchmark}
\subsection{Data Collection:}
To ensure \our{}'s comprehensiveness, various factors are considered. Data should originate from multiple cohorts, where each cohort is a unique country. For a representative collection of cancer types, datasets with multiple cancers, cancer subtypes, rare cancer types, and grades are considered. To accommodate distribution shifts of acquisition tools, images must originate from multiple scanners and microscopes, and tiles must be patched at different magnification levels. 
An overall description of the data is shown in Table~\ref{tab:dataCharacteristics}. 

\begin{table*}[h!]
\centering
\begin{adjustbox}{width=1\textwidth}
\begin{tabular}{lcccccccc}
\toprule
\bfseries Dataset& \bfseries Cancer Type & \bfseries Patches & \bfseries WSIs & \bfseries Dimensions & \bfseries Magnification & \bfseries Origin \\

\midrule
\rowcolor{brown!5}
CAMEL~\cite{CAMEL} & Colon & 15,403 & 177 & $1280\times1280$& 20X & China\\
\rowcolor{brown!5}
GlaS~\cite{glas} & Colorectal glands & 165 & 16 & $567-775\times430-522$ & 20X & N/A \\
\rowcolor{brown!5}
DiagestPath~\cite{Digestpath} & gastric/colorectal & 107,076 & N/A &$224\times224$ & N/A & N/A \\
\rowcolor{brown!5}
LC25000~\cite{LC25K} &  Colon, Lungs & 25000 & 150 & $768\times768$ & 60X & USA\\
\rowcolor{brown!5}
MHIST~\cite{mhist} & Colorectal & 3,152 & 279 & $224\times224$ & 20X &USA\\
\rowcolor{brown!5}
PanNuke~\cite{gamper2019pannuke,pannuke2} &  19 &  7,901 & 27724 &  $256\times256$ & 40X & USA+UK\\
\rowcolor{brown!5}
WSSS4LUAD~\cite{wsss4luad1,wsss4luad2} & Lung & 10,211 & 67 & $150-5606\times150-4371$ & 40X & China\\
\rowcolor{brown!5}
Breast IDC~\cite{idcKaggle} & Breast & 277,524 & 162 & $50\times50$ & 40X & N/A \\
\rowcolor{brown!5}
GasHisSDB~\cite{GasHisSDB} & Gastric  & 245,196 & N/A & $80\times80$,$120\times120$,$160\times160$ & N/A & N/A \\
\rowcolor{magenta!8}
CRC-ICM~\cite{crcIcm} & Colorectal & 1,744 & 136 & $1280-4140\times960-3096$ & 200X & Iran \\
\rowcolor{magenta!8}
DataBiox~\cite{databiox} & Breast  & 922 & 124 & $1276-4032\times956-3024$ & 4X,10X,20X,40X & Iran \\
\rowcolor{magenta!8}
PatchGastric~\cite{patchGastric} & Stomach & 262,777 & 991 & $300\times300$ & 20X & Japan\\
\rowcolor{magenta!8}
SICAPv2~\cite{sicap} & Prostate & 12,081 & 155 & $512\times512$ & 10X & Spain\\
\rowcolor{magenta!8}
Prostate Grading~\cite{prostateGradingHarvard} & Prostate & 641 & N/A & $3100\times3100$ & 40X & Switzerland \\
\rowcolor{magenta!8}
GlaS~\cite{glas} & Colorectal glands & 165 & 16 & $567-775\times430-522$ & 20X & N/A \\
\rowcolor{blue!8}
CRC-ICM~\cite{crcIcm} & Colorectal & 1,744 & 136 & $1280-4140\times960-3096$ & 200X & Iran \\
\rowcolor{blue!8}
BRACS~\cite{BRACS}  & Breast & 4,539 & 547 & $127-17611\times137-13462$& 40X & Italy\\
\rowcolor{blue!8}
BCNB~\cite{bcnb}  & Breast & 76,578 & 1058 & $256\times256$ & - & China\\
\rowcolor{blue!8}
BreakHist~\cite{spanhol2015dataset} & Breast & 7,909 & 82 & $700\times460$ & 40X,100X,200X,400X & Brazil\\
\rowcolor{blue!8}
MSI vs. MSS (FFPE)~\cite{msivsmss} & Stomach,Colon & 411,890 & N/A & $224\times244$ & 10X & USA \\
\rowcolor{blue!8}
BACH~\cite{bach} & Breast & 400 & 10 & $2048\times1536$ & 10X & Portugal\\
\rowcolor{blue!8}
SkinTumor~\cite{skinCancer} & Skin & 129,369 & 378 & $395\times395$ & 400X & Germany \\
\rowcolor{blue!8}
PatchGastric~\cite{patchGastric} & Stomach & 262,777 & 991 & $300\times300$ & 20X & Japan\\
\rowcolor{blue!8}
TCGA-Uniform~\cite{tcgaUniform} & 32 & 1,608,060 & 7,951 & $256\times256$ & 5X-10X & USA\\
\rowcolor{blue!8}
MPNv2~\cite{mpn} &  blood & 300 & N/A & $1297-1297\times22160$ & N/A & Malaysia \\
\rowcolor{yellow!8}
CRC-100K~\cite{Kather2018_jx} &  Colorectal,Stomach & 207180  & 136 & $224\times224$ & 10X & Germany\\
\rowcolor{yellow!8}
Kather-16~\cite{Kather2016_fm} &  Colorectal & 5000 & 10 & $224\times224$ & 20X & Germany\\
\rowcolor{yellow!8}
Osteosarcoma~\cite{sarcoma1,sarcoma2} & Bone & 1,144 & 4 & $1024\times1024$ & 10X & USA \\
\rowcolor{yellow!8}
BACH~\cite{bach}  & Breast  & 400 & 29 & $2048\times1563$ & 10X & Portugal\\
\rowcolor{yellow!8}
RenalCell~\cite{renalCell1} & Kidney & 64,553 & 497 & $300\times300$, $256\times256$ & 10X,20X & USA+Finland\\
\rowcolor{yellow!8}
CRC-TP~\cite{crcTp} & Colorectal & 280,000 & 20 & $150\times150$ & 20X &  UK\\
\rowcolor{yellow!8}
Chaoyang~\cite{chaoyang} & Colon & 5,160 & N/A & $512\times512$ & 20X & USA \\
\rowcolor{yellow!8}
SkinCancer~\cite{skinCancer} & Skin & 129,369 & 378 & $395\times395$ & 400X & Germany \\
\rowcolor{orange!8}
PCam~\cite{pcam} & Lymph nodes & 327,680 & 400 & $96\times96$ & 10X & Netherlands\\
\rowcolor{orange!8}
BCNB~\cite{bcnb} & Breast & 76,578 & 1058 & $256\times256$ & - & China\\
\rowcolor{green!8}
TCGA-TIL~\cite{tcgaTIL1,tcgaTIL2,tcgaTIL3} & Multiple & 304,097 & N/A &$100\times100$ & 10X & USA \\
\rowcolor{green!8}
RenalCell~\cite{renalCell1} & Kidney & 64,553 & 497 & $300\times300$, $256\times256$ & 10X,20X & USA+Finland\\
\rowcolor{pink!25}
MSI vs. MSS (FFPE)~\cite{msivsmss} & Stomach,Colon & 411,890 & N/A & $224\times244$ & 10X & USA \\
\rowcolor{pink!25}
MSI vs. MSS~\cite{snapFrozen} (Snap Frozen) & Stomach & 218,578 & N/A & $224\times244$ & 10X & USA \\

\bottomrule
\rowcolor{gray!15}
Overall Numbers & 32 & 5,358,942 & 41,508 & $96-17611\times96-13462$ & 10 levels & 14 \\

\bottomrule
\end{tabular}
\end{adjustbox}
\caption{Benchmark data description including 32 cancer types, 5.3M patches tiled from 41K WSIs across 14 different countries. Further information regarding the organs, patch dimensions, and magnification levels, are also listed. Datasets are color-coded for different clinical tasks, including cancer detection(\textit{gray}), cancer grading (\textit{magenta}), cancer subtyping (\textit{blue}), tissue phenotyping (\textit{yellow}), metastasis detection (\textit{orange}), TIL detection (\textit{green}), and MSI detection (\textit{pink}).}
\label{tab:dataCharacteristics}
\end{table*}

\subsection{Caption Generation:} 
\begin{figure}[t!]
\centering
\includegraphics[width=\textwidth]{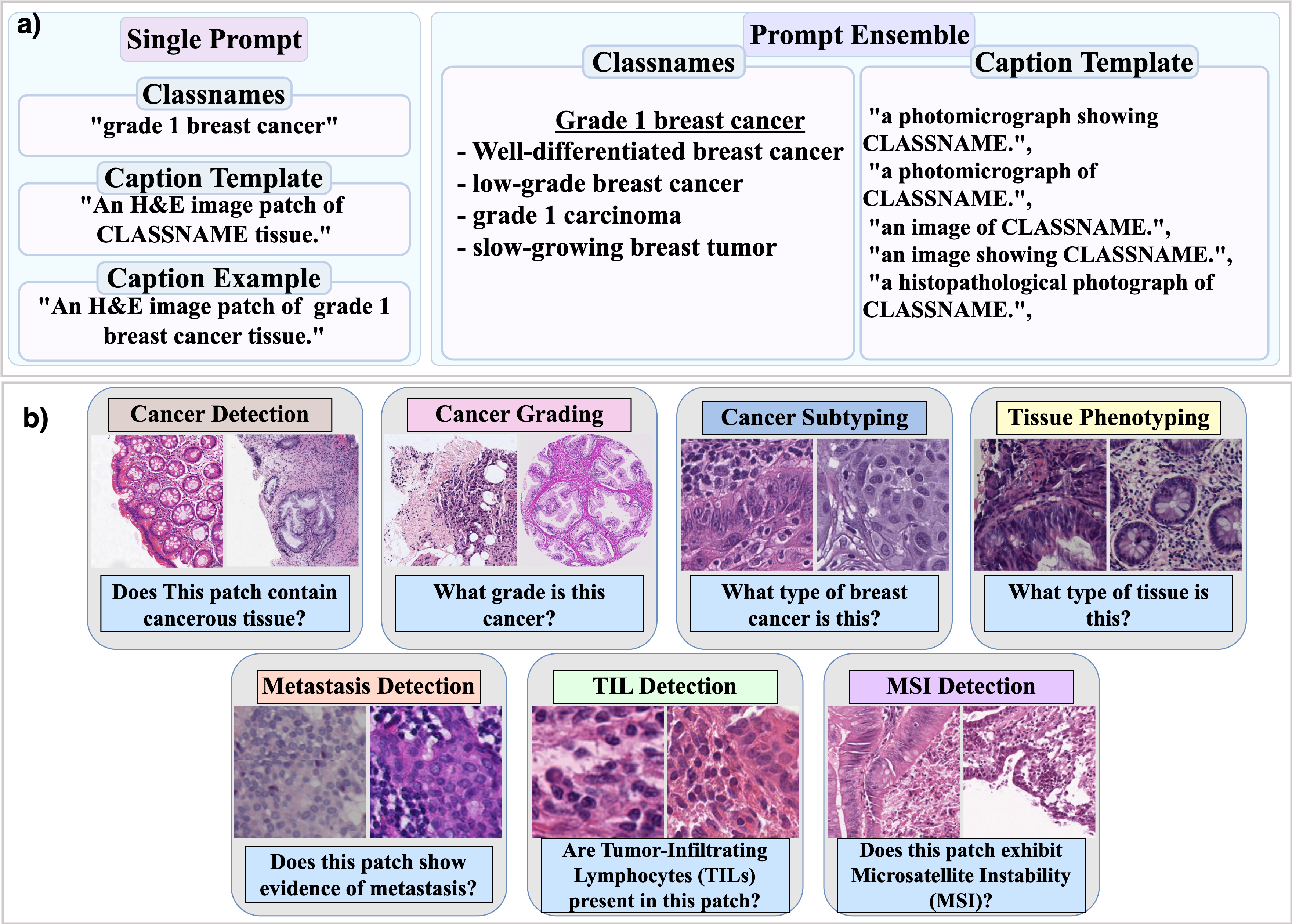}
\caption{(a) Single and ensemble captions generated. Single caption is a fixed template incorporating the classname, while ensemble captions provide higher variability and information in the template and classname description(b) Different clinical tasks that simulate real-life actions performed by doctors.}
\label{captionGen}
\end{figure}

Two caption types are created for each image in the benchmark: a single caption and an ensemble of captions. The single caption uses a template (similar to~\cite{plip}) incorporating the class name, while the ensemble assigns four different synonyms for each classname generated using LLMs~\cite{chatGpt} and incorporates them into 22 different templates following the ensemble templates in~\cite{conch}. An example of the classnames is displayed in Tanle~\ref{tab:captionEnsemble}. Figure~\ref{captionGen}\textbf{(a)} displays examples of both caption types.

\subsection{Experimental Setup}
\subsubsection{Clinical Tasks:} For clinical and experimental consistency, data information from dataset description, provided metadata, and classes, are used to categorize datasets into representative 
 real clinical tasks scenarios. 7 main clinical tasks are addressed:
(1)\textit{Cancer Detection:} binary classification of the presence of cancerous tissues. (2)\textit{Cancer Grading:} multiclass classification of the cancer grade, representing its aggressiveness. (3)\textit{Cancer Subtyping:} multiclass classification of cancer subtypes. (4)\textit{Tissue Phenotyping:} multiclass classification of different tissue types. (5)\textit{Metastasis Detection:} binary classification of the presence of metastasis. (6)\textit{TIL Detection:} binary classification of the presence of Tumor Infiltrating Lymphocytes, and (7)\textit{MSI Detection:} binary classification of the presence of Microsatellite instability (MSI). These tasks simulate real-life tasks performed by doctors to assess cancers and create a treatment plan accordingly, which will help better understand the models' performance. The tasks are summarized in Figure~\ref{captionGen}\textbf{(b)}.

 \noindent Default augmentations are resize, center crop, and normalize. The dimensions of the processed images are $224\times224$ or $448\times448$, depending on the model's input size. Some datasets were used in multiple tasks due to the presence of multiple groundtruths for the same image (E.g. same dataset has multiple cancer subtypes and other tissue types, like SkinCancer dataset). To accommodate the large imbalance of datasets in performance evaluation, multiple metrics are used including balanced accuracy, F1-score, precision, and Matthews correlation coefficient (MCC). Reliability plots and estimated calibration error (ECE) are also used to assess model uncertainty and calibration.

\section{Experiments \& Results}


\begin{figure}[htbp]
\includegraphics[width=\textwidth]{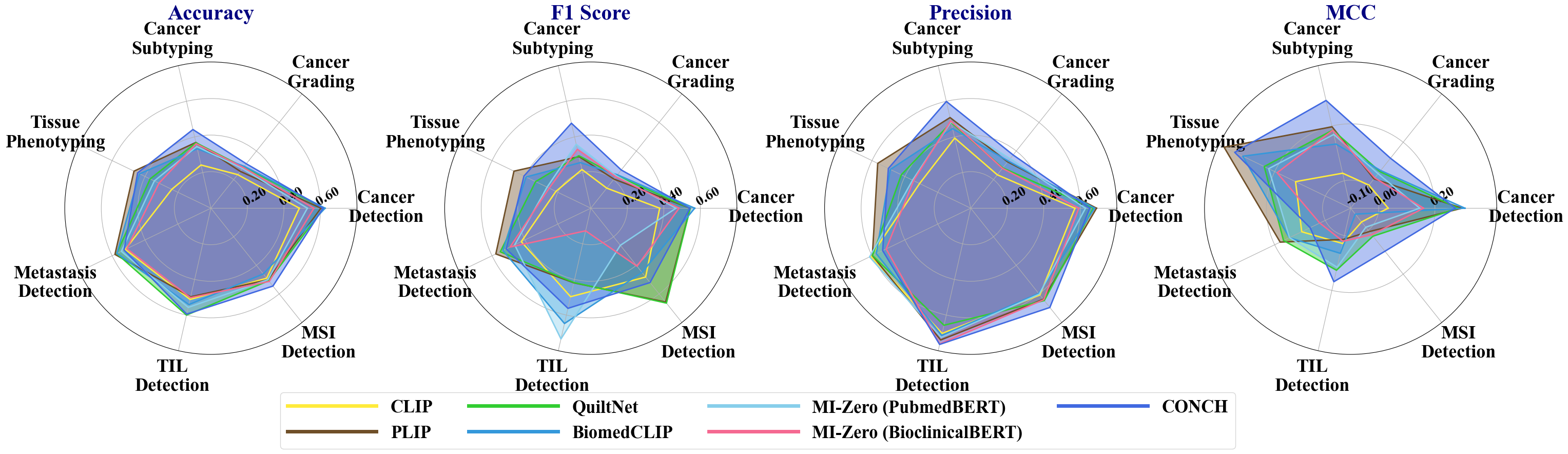}
\caption{Models performance in single caption scenario. Metrics displayed include balanced accuracy, F1-score, precision and MCC.}
\label{singlPerformance}
\end{figure}

\begin{figure}[!t]
\includegraphics[width=\textwidth]{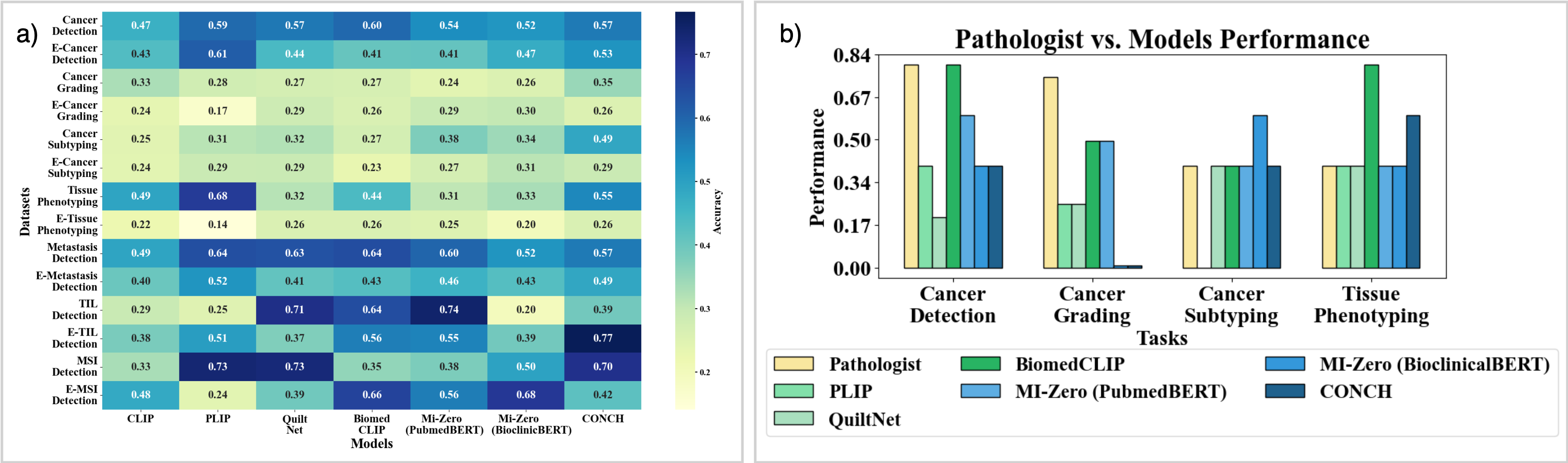}
\caption{(a) Models performance in ensemble captions scenario. The heatmap displays the balanced accuracies of models across different tasks, where "E-" signifies ensemble. (b) Comparison between an expert's and the VLMs' performance }
\label{ensPath}
\end{figure}

\subsection{Performance Analysis of models on clinical tasks} 
Figure~\ref{singlPerformance} highlights the average performance of models on different tasks in single caption scenario. Domain-specific models demonstrate better balanced accuracies than general (BiomedCLIP) and nonmedical (CLIP) VLMs. However, most VLMs demonstrate better precision in binary detection tasks, which is affected by data imbalance. QuiltNet archives the highest F1 score in MSI detection, PLIP in tissue phenotyping and metastasis detection, MI-Zero's PubmedBERT in TIL detection, and CONCH in cancer grading and subtyping. However, if we consider all the metrics, including MCC we see that CONCH relatively has the best performance.

\subsubsection{Caption Change Effect}
To test the effect of caption content on the performance of models, 10 different captions were used to evaluate how
each model performs on the same dataset when provided
with different textual descriptions, including details about
the organ of interest, magnification, and other class-
specific information. The models' performances on each of the captions are listed in Table~\ref{tab:captionVariation}. The results reveal that models are highly
sensitive to caption modifications, with balanced accu-
racy fluctuating significantly across captions, with fluctuations in balanced accuracy minimally observed in QuiltNet with 6.96\% difference in accuracy and the highly observed in CONCH reaching 26\%. These findings highlight the critical role of caption or prompt design
in determining model performance.

\begin{table}[ht!]\centering
    \begin{adjustbox}{width=1\textwidth}
    \begin{tabular}{lcccccccc}
    \toprule
     \multirow{2}{*}{\bfseries Caption} & \multirow{2}{*}{CLIP} & \multirow{2}{*}{PLIP} & \multirow{2}{*}{QuiltNet} & \multirow{2}{*}{BiomedCLIP} & MI-Zero & MI-Zero & \multirow{2}{*}{CONCH} \\
      & &  & & & (PubmedBERT) & (BioclinicalBERT) &  \\
    \bottomrule
    a photomicrograph showing adipose tissue. &  0.1906 & 0.5982 & 0.4659 & 0.5058 & 0.4281 & 0.4016 & \textbf{0.6651}\\
    
    an image of adipose tissue. & 0.2083 & 0.5796 & 0.5049 & 0.3683 & 0.4438 & 0.401 & 0.5014\\
    
    a histopathological photograph of adipose tissue. & 0.2338 & \textbf{0.6508} & 0.4607 & \textbf{0.5338} & \textbf{0.4661} & 0.4453 & 0.4197\\
    
    presence of adipose tissue. & 0.1584 & \underline{0.5078} & 0.4774 & \underline{0.3429} & 0.4417 & 0.4114 & \underline{0.3975} \\
    
    adipose, H\&E stain. &  0.2482 & 0.5612 & \underline{0.4488} & 0.5007 & 0.3312 & \textbf{0.5263} & 0.5992\\
    
    a colorectal photomicrograph showing adipose tissue. & 0.1925 & 0.5263 & 0.5083 & 0.5051 & 0.3595 & 0.4990 & 0.4819\\
    
    a photomicrograph at 10x showing adipose tissue. & 0.1946 & 0.5974 & 0.4812 & 0.4711 & 0.3684 & 0.3767 & 0.6132 \\
    
    an H\&E stained image showing adipose tissue. & \underline{0.1497} & 0.5627 & \textbf{0.5184} & 0.4259 & 0.4516 & 0.5109 & 0.6435\\
    
    adipose H\&E stain at 10x. &  \textbf{0.2622} & 0.5912 & 0.4875 & 0.4782 & 0.3151 & 0.5018 & 0.569\\
    
    colorectal adipose H\&E stain at 10x. & 0.1789 & 0.5415 & 0.4743 & 0.4743 & \underline{0.2391} & \underline{0.3698} & 0.4189\\
    
    \end{tabular}
    \end{adjustbox}
    \caption{Different captions used to test the effect of added/changed text input on the overall performance of the models.}
    \label{tab:captionVariation}
    \end{table}

\subsubsection{Caption Ensemble:} In the caption ensemble scenario, the performance fluctuates with large drops or gains in accuracy, as seen in Figure~\ref{ensPath}(\textbf{a}). CONCH's performance drops by almost 30\% in tissue phenotyping but is enhanced by 38\% in TIL detection. The same trend of large fluctuations are observed in other models but are less severe in CONCH.

\subsection{Statistical Significance:}
For statistical significance, \textbf{Wilcoxon test}, a nonparametric signed-rank statistical test, is conducted in both single and ensemble scenarios. In single caption scenario, medical VLMs show statistically significant performance when compared to CLIP, with $p-value\leq0.05$, but among the medical VLMS, MI-Zero variants have statistically significant results when compared to QuiltNet with $p-values$ 0.0156 and 0.0469, while CONCH's significant differences are only evident when compared to MI-Zero's results with 0.0312 and 0.0156. Caption ensemble results, however, are different. Statistical significance is evident when comparing CONCH to CLIP, PLIP and QuiltNet and MI-Zero's PubmedBERT with $p-values$ of 0.0156 for the first three and 0.0496. Results of the Wilcoxon test can be observed in Table~\ref{tab:wilocxonTest}.

\begin{table}[t!]\centering
    \begin{adjustbox}{width=1\columnwidth}
    \begin{tabular}{lcccccc}
    \toprule
     \bfseries Model 1 & \bfseries Model 2 & \multicolumn{2}{c}{Single} & \multicolumn{2}{c}{Ensemble}\\
     \cmidrule(lr){3-4} \cmidrule(lr){5-6}
     & & \bfseries Stats & \bfseries $p-value$ & \bfseries Stats & \bfseries $p-value$ \\
    \bottomrule
      CLIP & PLIP & 2.0000 & \textbf{0.0469} & 14.0000 & 1.0000\\ 
      CLIP & quilt & 0.0000 & \textbf{0.0156} & 2.0000 & \textbf{0.0469} \\
      CLIP & BiomedCLIP & 2.0000 & \textbf{0.0469} & 5.0000 & 0.1562\\
      CLIP & MI-Zero(PubMedBERT) & 0.0000 & \textbf{0.0156} & 14.0000 & 1.0000\\
      CLIP & MI-Zero(BioclinicalBERT) & 3.0000 & 0.0781 & 10.0000 & 0.5781 \\
      CLIP & CONCH & 0.0000 & \textbf{0.0156} & 0.0000 & \textbf{0.0156}\\
       
      \midrule
      PLIP & QuiltNet & 10.0000 & 0.5781 &  10.0000 & 0.5781\\
      PLIP & BiomedCLIP & 13.0000 & 0.9375 &  9.0000 & 0.4688\\
      PLIP & MI-Zero(PubMedBERT) & 7.0000 & 0.2969 &  14.0000 & 1.0000\\
      PLIP & MI-Zero(BioclinicalBERT) & 8.0000 & 0.3750 & 11.0000 & 0.6875\\
      PLIP & CONCH & 7.0000 & 0.2969 & 0.0000 & \textbf{0.0156}\\

      \midrule
      QuiltNet & BiomedCLIP & 12.0000 & 0.8125 &  10.0000 & 0.5781\\
      QuiltNet & MI-Zero(PubMedBERT) & 0.0000 & \textbf{0.0156} & 10.0000 & 0.5781\\
      QuiltNet & MI-Zero(BioclinicalBERT) & 2.0000 & \textbf{0.0469} & 12.0000 & 0.8125\\
      QuiltNet & CONCH & 8.0000 & 0.3750 & 0.0000 & \textbf{0.0156}\\
      \midrule
      BiomedCLIP & MI-Zero(PubMedBERT) & 9.0000 & 0.4688 &  12.0000 & 0.8125\\
      BiomedCLIP & MI-Zero(BioclinicalBERT) & 6.0000 & 0.2188 &  12.0000 & 0.8125\\
      BiomedCLIP & CONCH & 5.0000 & 0.1562 &  9.0000 & 0.4688\\
      \midrule
      MI-Zero(PubMedBERT) & MI-Zero(BioclinicalBERT) & 12.0000 & 0.8125 &  8.0000 & 0.3750\\
      MI-Zero(PubMedBERT) & CONCH & 1.0000 & \textbf{0.0312} & 2.0000 & \textbf{0.0469}\\
      \midrule
      MI-Zero(BioclinicalBERT) & CONCH & 0.0000 & \textbf{0.0156} & 3.0000 & \textbf{0.0781}\\
    
    \end{tabular}
    \end{adjustbox}
    \caption{Results of paired Wilcoxon Signed-Rank tests between all models. P-values less than 0.05 are marked in boldface.}
    \label{tab:wilocxonTest}
    \end{table}

\begin{figure}[t!]
\centerline{\includegraphics[width=\textwidth]{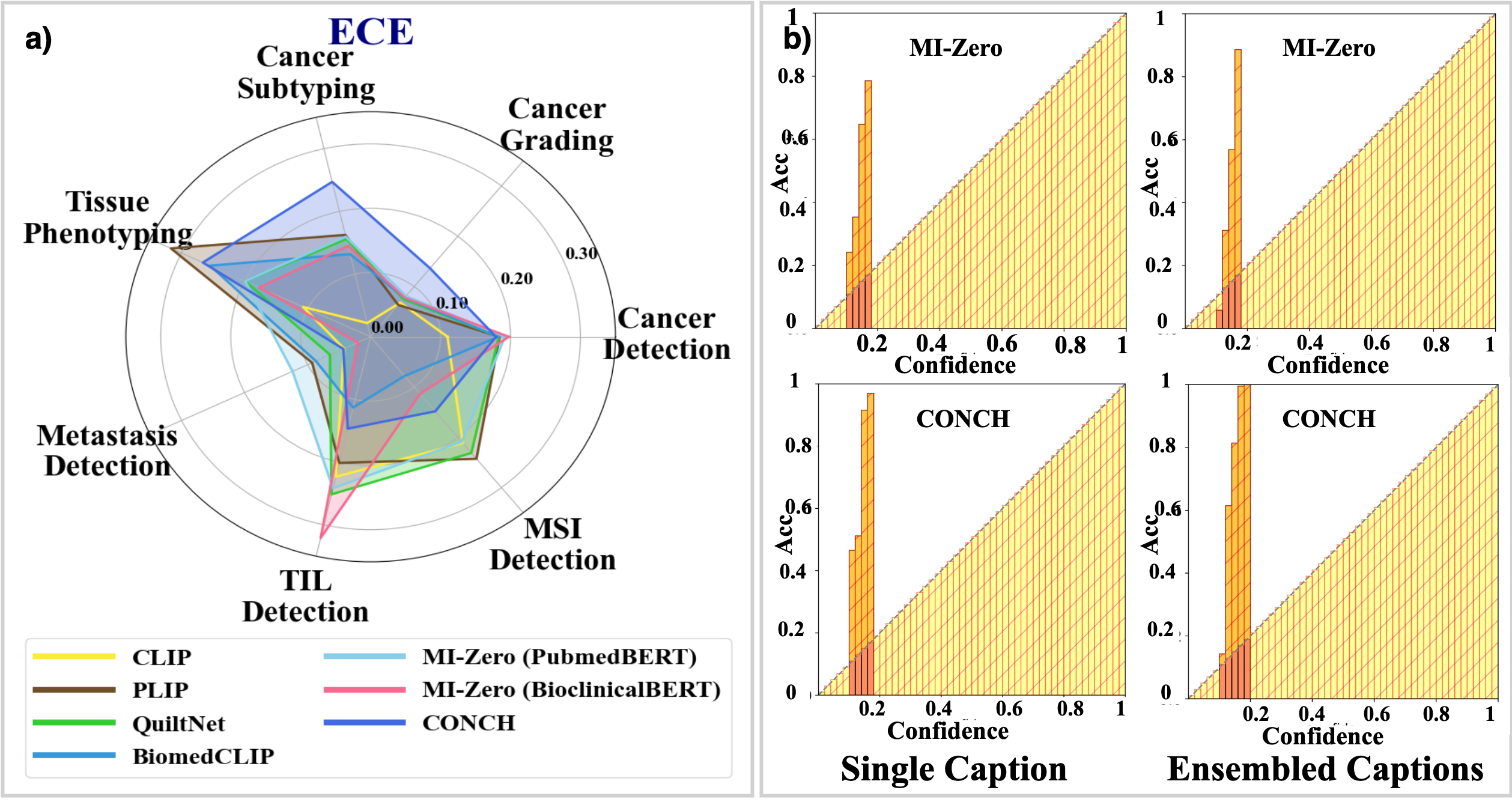}}
\caption{Measure of uncertainty in models' predictions. Subfigure \textbf{(a)} displays models' ECE values in a single prompt setting. Subfigure \textbf{(b)} displays an example of reliability plots of two models: BioclinicalBERT in the top row and CONCH in the bottom row. Each model's ECE is reported at the top of the graph.}
\label{eceSingle}
\end{figure}

\subsection{Model Calibration:} 
 As per Figure~\ref{eceSingle}(\textbf{a}), all models exhibit high ECE values across tasks. A general trend of the highest ECE values in tissue phenotyping, followed by TIL detection, MSI detection, and cancer grading is observed. Even CONCH, with its enhanced performance, exhibits high ECE values across most tasks. For further investigation, we study reliability plots of two models in single and ensemble caption scenarios on CRC-100K dataset; CONCH and MI-Zero's BioclinicalBERT seen in Figure~\ref{eceSingle}\textbf{(b)}. All graphs show very low confidence with high ECE values. BioclinicalBERT's ECE improves from 0.36 in single caption to 0.31, while CONCH's ECE increases from 0.53 to 0.61, even though it achieved an accuracy of 76\% on the CRC-100K dataset in the ensemble mode. \\

\begin{figure}[t!]
\centerline{\includegraphics[width=0.8\textwidth]{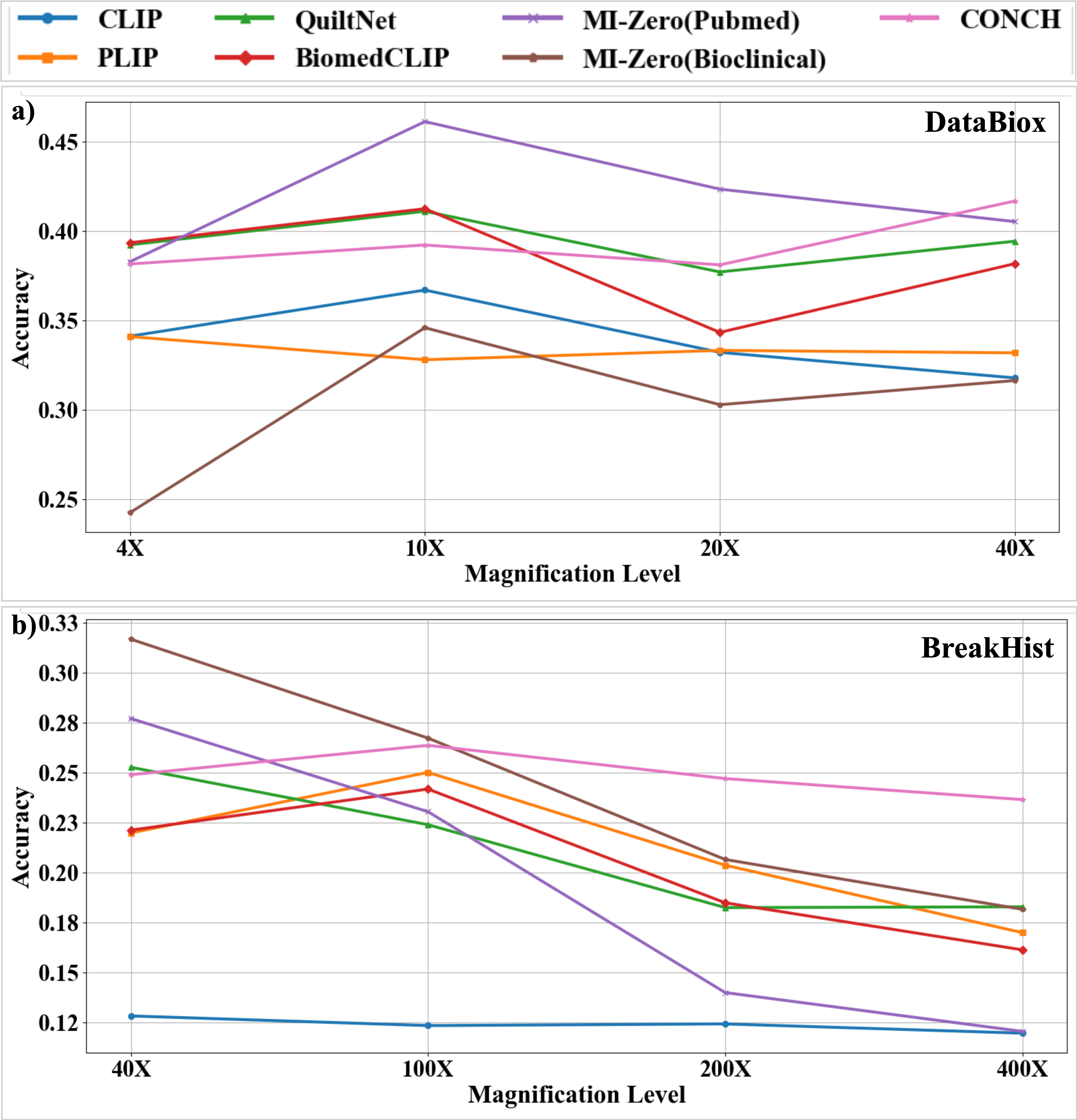}}
\caption{Impact of magnification on model performance across tasks. The top plot shows performance in cancer grading at various magnifications, while the bottom plot presents performance in cancer subtyping over different magnification ranges.}
\label{magnification}
\end{figure}

\subsection{Effect of Magnification}
Experiments on the BreakHist and DataBiox datasets indicate that magnification directly influences model performance. Although all images are resized uniformly, variations in magnification level determine the type of information fed to the model from cellular- to tissue-level information, each with different significance. Figure~\ref{magnification} presents balanced accuracies of models on(\textbf{a})  BreakHist with higher magnification levels (40X to 400X), and (\textbf{b}) on DataBiox, with lower magnification levels (4X to 40X). Most models exhibit better performance on 10X and 40X for cancer grading (DataBiox figure (\textbf{a}), while for cancer subtyping, higher magnifications are accompanied by a decrease in balanced accuracy. Based on the opinion of an expert pathologist, a 40X magnification is usually the most convenient for cancer grading, but the models tend perform better on 10X.

\begin{figure}[t!]
\centerline{\includegraphics[width=0.8\textwidth]{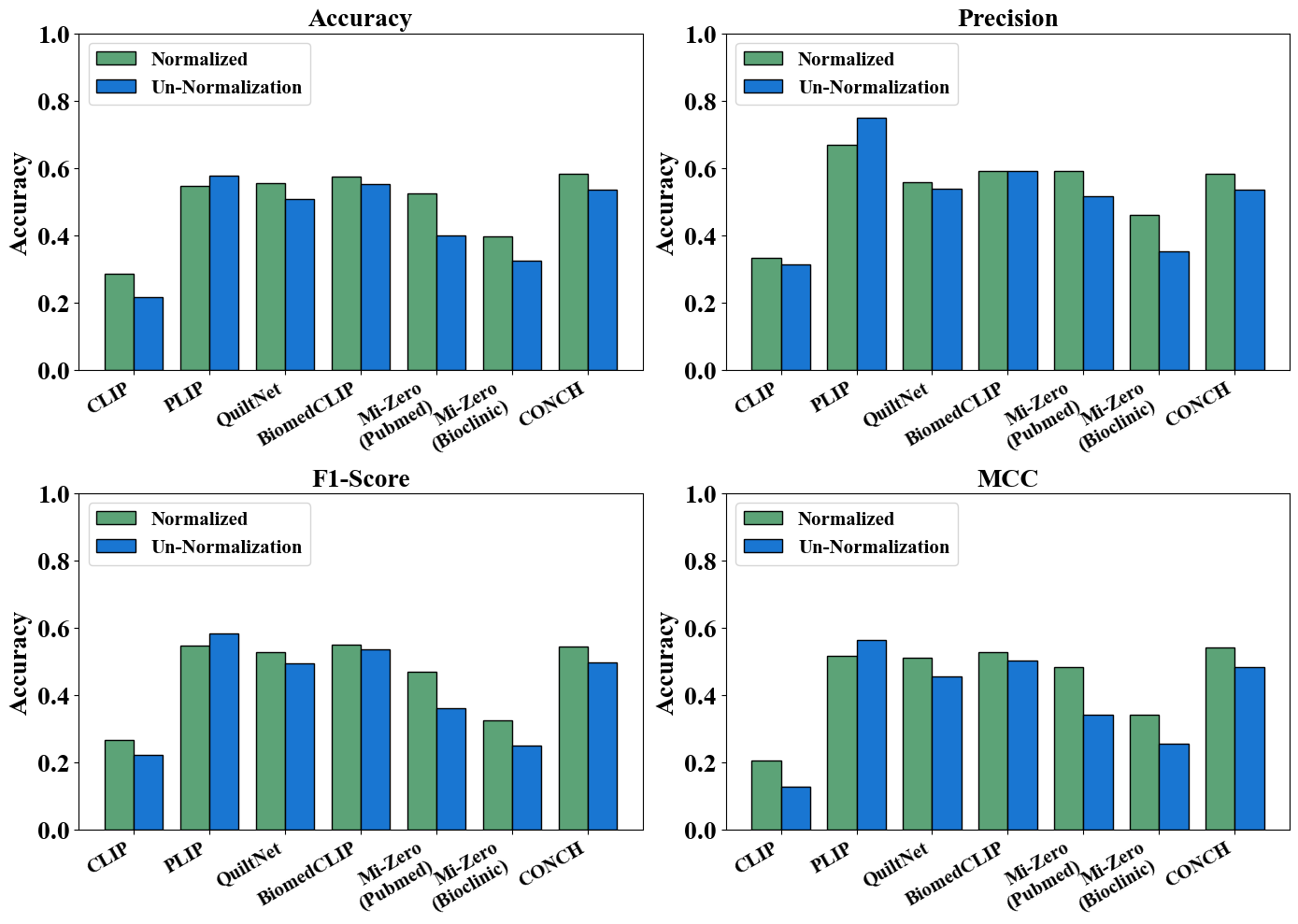}}
\caption{Effect of stain normalization on model performance.comparing patches from same WSIs with and without Macenko normalization. Dataset used is CRC-100K}
\label{normNonorm}
\end{figure}

\subsection{Effect of Normalization}
Tissue preparation usually follows multiple steps, each of which may affect the resulting stain~\cite{stainEffect}, therefore stain normalization is one of the important steps in histopathology image preprocessing~\cite{stain1,stain2}. To better understand its effect on the models' performances, we utilize two versions of the CRC-100K dataset, both of which were patched from the same WSIs, but one group was subjected to Macenko~\cite{macenko} normalization, and another with unnormalized patches. We sample a group of 2000 images per class and create a test set of randomly chosen 18000 images from each of the datasets and pass them through the models. Figure~\ref{normNonorm} visualizes the results, demonstrating a general drop in model performance in the absence of stain normalization, ranging between 2\% for BiomedCLIP up to 12\% for MI-Zero's PubmedBERT. This decline is expected, as the majority of publicly available datasets, as well as those used for training, are typically stain-normalized using methods like Macenko normalization. Consequently, unnormalized images negatively impact the performance of models trained on normalized data.

\subsection{Adversarial Corruption:}
\subsubsection{Visual Corruption:}
\begin{figure*}[t!]
\centerline{\includegraphics[width=\textwidth]{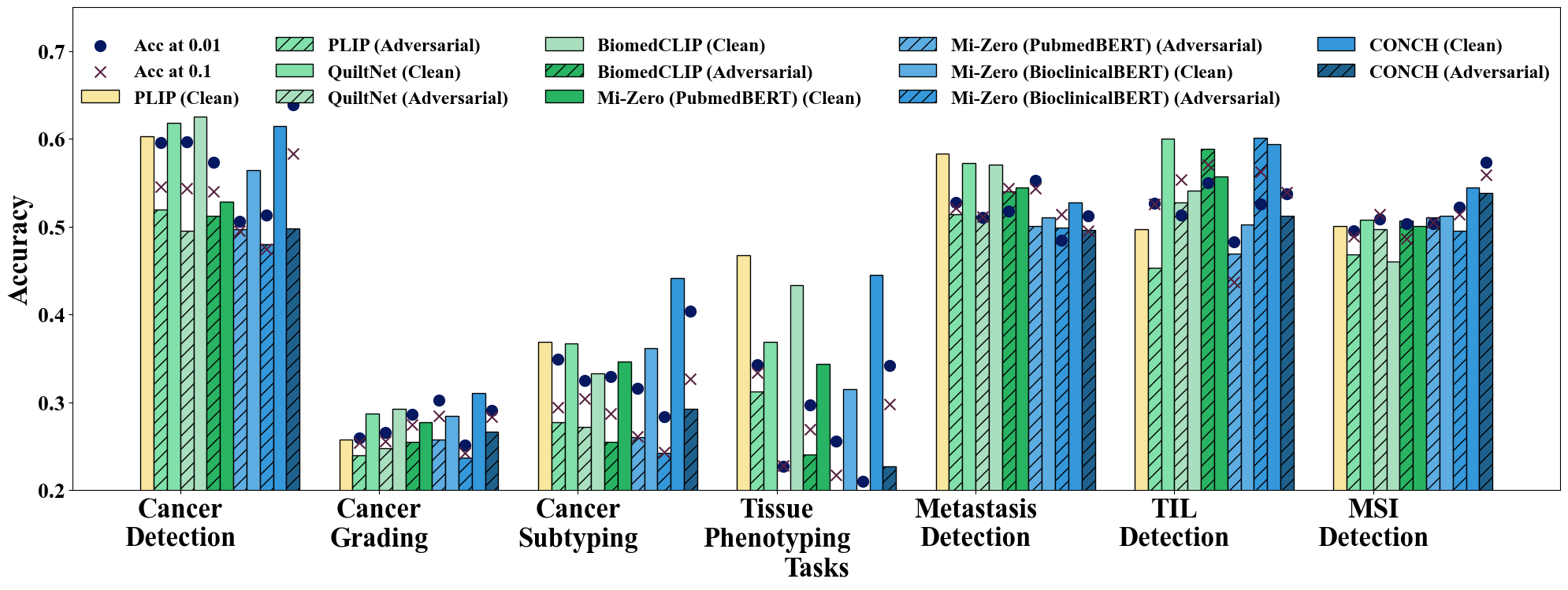}}
\caption{Blanaced accuracies of models across various tasks under 3 types of perturbation strengths. The navy dots represent accuracies of models at $\epsilon=0.01$, while dark red crosses represent accuracies at at $\epsilon=0.1$. The bars represent accuracies at $\epsilon=0.3$.}
\label{adverseResults}
\end{figure*}

To test the models' robustness, they are subjected to FGSM attacks at three perturbation levels: $\epsilon=[0.01,0.1,0.3]$. The results are displayed in Figure~\ref{adverseResults}. At the lowest perturbation level, the overall performances of the models are minimally affected. At $\epsilon=0.1$, more drastic drops in the accuracies are observed, reaching up to 14\%. With $\epsilon=0.3$, all models are greatly affected by the attack, with accuracy drops of up to 20\%.  Models trained on online data, like PLIP, QuiltNet and BiomedCLIP demonstrate better robustness even with higher perturbations, CONCH shows drops in accuracy at larger perturbations. An important observation is that models' accuracies dropped more significantly in tasks with more classes like cancer subtyping and tissue phenotyping, compared to those with binary or few-class classification.

\subsubsection{Textual Corruption}

\begin{table*}[h!]\centering
\begin{adjustbox}{width=1\textwidth}
\begin{tabular}{l|c|ccccccc}
\toprule
\multirow{2}{*}{\bfseries Dataset} & \multirow{2}{*}{\bfseries Corruption} & \multirow{2}{*}{\bfseries CLIP} & \multirow{2}{*}{\bfseries PLIP} & \multirow{2}{*}{\bfseries QuiltNet} & \multirow{2}{*}{\bfseries BiomedCLIP} & \bfseries MI-Zero & \bfseries MI-Zero & \multirow{2}{*}{\bfseries CONCH}\\
& & & & & &PubmedBERT&BioclinicalBERT&\\
\toprule
\multirow{4}{*}{\bfseries GasHisSDB} & Original & \underline{0.4732} & \underline{0.5}&0.5274 & 0.5366&  0.5223&\underline{0.4904} &0.625 \\
 & SWAP &0.5082 &0.5 & \underline{0.5241}&0.4979 &\underline{0.4957} &0.5513 &0.5117\\
 & Replace & 0.5374& 0.5739& 0.6563& 0.4997&0.5459 & 0.5016&\underline{0.4989} \\
 & Remove &0.5034 &0.574 &0.5891 & \underline{0.4745}& 0.5814& 0.502& 0.6229\\
 \midrule
\multirow{4}{*}{\bfseries Breast IDC} & Original &0.5528 &0.7532 & 0.5295&0.5835 &0.5196 &0.797 & 0.5434\\
 & SWAP & 0.5398 &\underline{0.3968} &\underline{0.452} &0.6907 &\underline{0.2645} &\underline{0.5017} & 0.7535\\
 & Replace & 0.5177& 0.4734& 0.5496&0.5471 & 0.7408& 0.5092&\underline{0.4979} \\
 & Remove &\underline{0.4863} &0.5646&0.4996& \underline{0.4989}&0.3348 &0.5504 & 0.5385\\
 \midrule
\multirow{4}{*}{\bfseries DataBiox} & Original & 0.3292& 0.3332&0.3952 &0.3893 &0.4167 &\underline{0.3058}&0.3988 \\
 & SWAP &0.3163 &0.3016&0.3395 &\underline{0.3834}&\underline{0.3082} & 0.3781& 0.3472\\
 & Replace & 0.3333& \underline{0.2905}& \underline{0.2922}& 0.3921&0.3812 & 0.3171&0.37 \\
 & Remove & \underline{0.2193}&0.3627 &0.3018 &0.4434 &0.4434 & 0.3325&\underline{0.3467} \\
 \midrule
\multirow{4}{*}{\bfseries LC25000-lung} & Original & 0.37507& 0.8096&0.7528 &0.4484 &0.3989& 0.6126&0.8535 \\
 & SWAP &0.333 & \underline{0.1538}&\underline{0.4779} & 0.333& 0.2899& \underline{0.3789}&0.6478 \\
 & Replace & \underline{0.2722}& 0.3213& 0.6423&\underline{0.2893} & \underline{0.2588}& \underline{0.3789}&\underline{0.3912} \\
 & Remove &0.3327 &0.1553 &0.5665 & 0.3166& 0.2815&0.4148 & 0.7399\\
 \midrule
\multirow{4}{*}{\bfseries CRC-100K} & Original &0.1586 & 0.5751& 0.4614&0.5529 &0.4985 &0.5104 &0.6125 \\
 & SWAP &0.1989 & 0.3862&0.3422 &0.2831 &0.3537 & 0.3804& 0.2753\\
 & Replace & 0.2166& \underline{0.3449}&\underline{0.1812}&\underline{0.2795} & \underline{0.1645}& \underline{0.3075}&\underline{0.2375} \\
 & Remove &\underline{0.1319} &0.4634 &0.4206 &0.4096 &0.4092& 0.3982&0.3948 \\
\bottomrule
\end{tabular}
\end{adjustbox}
\caption{Balanced Accuracy of models on different datasets from multiple tasks evaluated under different textual corruptions. Underlines represent the lowest performances.}
\label{tab:textCorruption}
\end{table*}

We also evaluated performance under various text corruptions, including letter swap, replace, and delete, with results presented in Table~\ref{tab:textCorruption}. The underscore highlights the lowest accuracy between the original and corrupted prompts. All models were negatively impacted by these corruptions. In detection tasks (first two datasets), letter swap had the most degrading effect, significantly impacting four of the seven models. In cancer grading (third dataset), all corruption types caused substantial performance drops, while in cancer subtyping, letter replace was the most detrimental. For phenotyping, letter replacing consistently resulted in the worst accuracy. None of the models demonstrated robust resistance to these corruptions, further emphasizing the susceptibility of VLMs to variations in textual input.

\subsection{Pathologist vs. Model Performance:}
For further validation, we conduct a comparison with a pathologist of almost 30 years of experience. In this setting, representative images are randomly selected from the datasets covering various clinical tasks, and are presented in a questionnaire format with the same classes and information provided to the models. The results, shown in Figure~\ref{ensPath}(\textbf{b}), reveal that BiomedCLIP and MI-Zero's PubmedBERT overall achieve a performance comparable to the pathologist while other models fail in some of tasks. On the other hand, in multiclass classification tasks most models perform well, with some having higher accuracy than the pathologist, while in binary and few-class classifications, the pathologist performs better than the models.

\section{Discussion \& Conclusion}
This study presents a holistic vision language benchmark constructed to simulate real-world challenging scenarios and provide a unified platform for assessing histopathology VLMs. The comprehensive analysis conducted on current models shows severe fluctuations in accuracy upon caption changes, the low confidence and high ECE values consistently observed across models, and tasks, and the lack of adversarial robustness, emphasized through the large drops in accuracy in both image and text modalities, as well as the models' sensitivity to magnification levels as well as stain normalization of images, demonstrate the models' high sensitivity to textual inputs, lack of proper calibration and vulnerability, raising concerns about the reliability of these models in clinical applications. From here, multiple future directions can be targeted, including training more robust text encoders, focusing on model calibration, and incorporating adversarial training recipes for models to be deployed.

\bibliographystyle{IEEEtran}
\bibliography{ref}

\begin{thebibliography}{10}
\providecommand{\url}[1]{#1}
\csname url@samestyle\endcsname
\providecommand{\newblock}{\relax}
\providecommand{\bibinfo}[2]{#2}
\providecommand{\BIBentrySTDinterwordspacing}{\spaceskip=0pt\relax}
\providecommand{\BIBentryALTinterwordstretchfactor}{4}
\providecommand{\BIBentryALTinterwordspacing}{\spaceskip=\fontdimen2\font plus
\BIBentryALTinterwordstretchfactor\fontdimen3\font minus \fontdimen4\font\relax}
\providecommand{\BIBforeignlanguage}[2]{{%
\expandafter\ifx\csname l@#1\endcsname\relax
\typeout{** WARNING: IEEEtran.bst: No hyphenation pattern has been}%
\typeout{** loaded for the language `#1'. Using the pattern for}%
\typeout{** the default language instead.}%
\else
\language=\csname l@#1\endcsname
\fi
#2}}
\providecommand{\BIBdecl}{\relax}
\BIBdecl

\bibitem{clip}
A.~Radford, J.~W. Kim, C.~Hallacy, A.~Ramesh, G.~Goh, S.~Agarwal, G.~Sastry, A.~Askell, P.~Mishkin, J.~Clark \emph{et~al.}, ``Learning transferable visual models from natural language supervision,'' in \emph{International conference on machine learning}.\hskip 1em plus 0.5em minus 0.4em\relax PMLR, 2021, pp. 8748--8763.

\bibitem{align}
C.~Jia, Y.~Yang, Y.~Xia, Y.-T. Chen, Z.~Parekh, H.~Pham, Q.~Le, Y.-H. Sung, Z.~Li, and T.~Duerig, ``Scaling up visual and vision-language representation learning with noisy text supervision,'' in \emph{International conference on machine learning}.\hskip 1em plus 0.5em minus 0.4em\relax PMLR, 2021, pp. 4904--4916.

\bibitem{biomedclip}
S.~Zhang, Y.~Xu, N.~Usuyama, H.~Xu, J.~Bagga, R.~Tinn, S.~Preston, R.~Rao, M.~Wei, N.~Valluri \emph{et~al.}, ``Biomedclip: a multimodal biomedical foundation model pretrained from fifteen million scientific image-text pairs,'' \emph{arXiv preprint arXiv:2303.00915}, 2023.

\bibitem{plip}
Z.~Huang, F.~Bianchi, M.~Yuksekgonul, T.~J. Montine, and J.~Zou, ``A visual--language foundation model for pathology image analysis using medical twitter,'' \emph{Nature Medicine}, pp. 1--10, 2023.

\bibitem{quilt}
W.~O. Ikezogwo, M.~S. Seyfioglu, F.~Ghezloo, D.~S.~C. Geva, F.~S. Mohammed, P.~K. Anand, R.~Krishna, and L.~Shapiro, ``Quilt-1m: One million image-text pairs for histopathology,'' \emph{arXiv preprint arXiv:2306.11207}, 2023.

\bibitem{conch}
M.~Y. Lu, B.~Chen, D.~F. Williamson, R.~J. Chen, I.~Liang, T.~Ding, G.~Jaume, I.~Odintsov, L.~P. Le, G.~Gerber \emph{et~al.}, ``A visual-language foundation model for computational pathology,'' \emph{Nature Medicine}, vol.~30, p. 863–874, 2024.

\bibitem{miZero}
M.~Y. Lu, B.~Chen, A.~Zhang, D.~F. Williamson, R.~J. Chen, T.~Ding, L.~P. Le, Y.-S. Chuang, and F.~Mahmood, ``Visual language pretrained multiple instance zero-shot transfer for histopathology images,'' in \emph{Proceedings of the IEEE/CVF conference on computer vision and pattern recognition}, 2023, pp. 19\,764--19\,775.

\bibitem{pcam}
B.~S. Veeling, J.~Linmans, J.~Winkens, T.~Cohen, and M.~Welling, ``Rotation equivariant cnns for digital pathology,'' in \emph{Medical Image Computing and Computer Assisted Intervention--MICCAI 2018: 21st International Conference, Granada, Spain, September 16-20, 2018, Proceedings, Part II 11}.\hskip 1em plus 0.5em minus 0.4em\relax Springer, 2018, pp. 210--218.

\bibitem{paip}
Y.~J. Kim, H.~Jang, K.~Lee, S.~Park, S.-G. Min, C.~Hong, J.~H. Park, K.~Lee, J.~Kim, W.~Hong \emph{et~al.}, ``Paip 2019: Liver cancer segmentation challenge,'' \emph{Medical image analysis}, vol.~67, p. 101854, 2021.

\bibitem{bach}
G.~Aresta, T.~Ara{\'u}jo, S.~Kwok, S.~S. Chennamsetty, M.~Safwan, V.~Alex, B.~Marami, M.~Prastawa, M.~Chan, M.~Donovan \emph{et~al.}, ``Bach: Grand challenge on breast cancer histology images,'' \emph{Medical image analysis}, vol.~56, pp. 122--139, 2019.

\bibitem{campanella2024clinical}
G.~Campanella, S.~Chen, R.~Verma, J.~Zeng, A.~Stock, M.~Croken, B.~Veremis, A.~Elmas, K.-l. Huang, R.~Kwan \emph{et~al.}, ``A clinical benchmark of public self-supervised pathology foundation models,'' \emph{arXiv preprint arXiv:2407.06508}, 2024.

\bibitem{breen2024histopathology}
J.~Breen, K.~Allen, K.~Zucker, L.~Godson, N.~M. Orsi, and N.~Ravikumar, ``Histopathology foundation models enable accurate ovarian cancer subtype classification,'' \emph{arXiv preprint arXiv:2405.09990}, 2024.

\bibitem{neidlinger2024benchmarking}
P.~Neidlinger, O.~S. El~Nahhas, H.~S. Muti, T.~Lenz, M.~Hoffmeister, H.~Brenner, M.~van Treeck, R.~Langer, B.~Dislich, H.~M. Behrens \emph{et~al.}, ``Benchmarking foundation models as feature extractors for weakly-supervised computational pathology,'' \emph{arXiv preprint arXiv:2408.15823}, 2024.

\bibitem{kang2023benchmarking}
M.~Kang, H.~Song, S.~Park, D.~Yoo, and S.~Pereira, ``Benchmarking self-supervised learning on diverse pathology datasets,'' in \emph{Proceedings of the IEEE/CVF Conference on Computer Vision and Pattern Recognition}, 2023, pp. 3344--3354.

\bibitem{CAMEL}
\BIBentryALTinterwordspacing
G.~Xu, Z.~Song, Z.~Sun, C.~Ku, Z.~Yang, C.~Liu, S.~Wang, J.~Ma, and W.~Xu, ``Camel: A weakly supervised learning framework for histopathology image segmentation,'' \emph{arXiv}, no. arXiv:1908.10555, Aug. 2019, arXiv:1908.10555 [cs, eess]. [Online]. Available: \url{http://arxiv.org/abs/1908.10555}
\BIBentrySTDinterwordspacing

\bibitem{glas}
\BIBentryALTinterwordspacing
K.~Sirinukunwattana, J.~P.~W. Pluim, H.~Chen, X.~Qi, P.-A. Heng, Y.~B. Guo, L.~Y. Wang, B.~J. Matuszewski, E.~Bruni, U.~Sanchez, A.~Böhm, O.~Ronneberger, B.~B. Cheikh, D.~Racoceanu, P.~Kainz, M.~Pfeiffer, M.~Urschler, D.~R.~J. Snead, and N.~M. Rajpoot, ``Gland segmentation in colon histology images: The glas challenge contest,'' \emph{arXiv}, no. arXiv:1603.00275, Sep. 2016. [Online]. Available: \url{http://arxiv.org/abs/1603.00275}
\BIBentrySTDinterwordspacing

\bibitem{Digestpath}
\BIBentryALTinterwordspacing
K.~Dataset, ``\BIBforeignlanguage{en}{Digestpath dataset}.'' [Online]. Available: \url{https://www.kaggle.com/datasets/mittalswathi/digestpath-dataset}
\BIBentrySTDinterwordspacing

\bibitem{LC25K}
A.~A. Borkowski, M.~M. Bui, L.~B. Thomas, C.~P. Wilson, L.~A. DeLand, and S.~M. Mastorides, ``Lung and colon cancer histopathological image dataset (lc25000),'' \emph{arXiv preprint arXiv:1912.12142}, 2019.

\bibitem{mhist}
\BIBentryALTinterwordspacing
J.~Wei, A.~Suriawinata, B.~Ren, X.~Liu, M.~Lisovsky, L.~Vaickus, C.~Brown, M.~Baker, N.~Tomita, L.~Torresani, J.~Wei, and S.~Hassanpour, \emph{\BIBforeignlanguage{en}{A Petri Dish for Histopathology Image Analysis}}, ser. Lecture Notes in Computer Science.\hskip 1em plus 0.5em minus 0.4em\relax Cham: Springer International Publishing, 2021, vol. 12721, p. 11–24. [Online]. Available: \url{https://link.springer.com/10.1007/978-3-030-77211-6_2}
\BIBentrySTDinterwordspacing

\bibitem{gamper2019pannuke}
J.~Gamper, N.~Alemi~Koohbanani, K.~Benet, A.~Khuram, and N.~Rajpoot, ``Pannuke: an open pan-cancer histology dataset for nuclei instance segmentation and classification,'' in \emph{Digital Pathology: 15th European Congress, ECDP 2019, Warwick, UK, April 10--13, 2019, Proceedings 15}.\hskip 1em plus 0.5em minus 0.4em\relax Springer, 2019, pp. 11--19.

\bibitem{pannuke2}
------, ``\BIBforeignlanguage{en}{Pannuke: An open pan-cancer histology dataset for nuclei instance segmentation and classification},'' in \emph{\BIBforeignlanguage{en}{Digital Pathology}}, C.~C. Reyes-Aldasoro, A.~Janowczyk, M.~Veta, P.~Bankhead, and K.~Sirinukunwattana, Eds.\hskip 1em plus 0.5em minus 0.4em\relax Cham: Springer International Publishing, 2019, p. 11–19.

\bibitem{wsss4luad1}
C.~Han, J.~Lin, J.~Mai, Y.~Wang, Q.~Zhang, B.~Zhao, X.~Chen, X.~Pan, Z.~Shi, Z.~Xu, S.~Yao, L.~Yan, H.~Lin, X.~Huang, C.~Liang, G.~Han, and Z.~Liu, ``Multi-layer pseudo-supervision for histopathology tissue semantic segmentation using patch-level classification labels,'' \emph{Medical Image Analysis}, p. 102487, 2022.

\bibitem{wsss4luad2}
C.~Han, X.~Pan, L.~Yan, H.~Lin, B.~Li, S.~Yao, S.~Lv, Z.~Shi, J.~Mai, J.~Lin \emph{et~al.}, ``Wsss4luad: Grand challenge on weakly-supervised tissue semantic segmentation for lung adenocarcinoma,'' \emph{arXiv preprint arXiv:2204.06455}, 2022.

\bibitem{idcKaggle}
A.~Janowczyk and A.~Madabhushi, ``Deep learning for digital pathology image analysis: A comprehensive tutorial with selected use cases,'' \emph{Journal of pathology informatics}, vol.~7, no.~1, p.~29, 2016.

\bibitem{GasHisSDB}
\BIBentryALTinterwordspacing
C.~Li. [Online]. Available: \url{https://gitee.com/neuhwm/GasHisSDB}
\BIBentrySTDinterwordspacing

\bibitem{crcIcm}
\BIBentryALTinterwordspacing
Z.~Mokhtari, E.~Amjadi, H.~Bolhasani, Z.~Faghih, A.~Dehghanian, and M.~Rezaei, ``Crc-icm: Colorectal cancer immune cell markers pattern dataset,'' 2023. [Online]. Available: \url{https://arxiv.org/abs/2308.10033}
\BIBentrySTDinterwordspacing

\bibitem{databiox}
\BIBentryALTinterwordspacing
H.~Bolhasani, E.~Amjadi, M.~Tabatabaeian, and S.~J. Jassbi, ``A histopathological image dataset for grading breast invasive ductal carcinomas,'' \emph{Informatics in Medicine Unlocked}, vol.~19, p. 100341, 2020. [Online]. Available: \url{https://www.sciencedirect.com/science/article/pii/S2352914820300757}
\BIBentrySTDinterwordspacing

\bibitem{patchGastric}
\BIBentryALTinterwordspacing
M.~Tsuneki and F.~Kanavati, ``Inference of captions from histopathological patches,'' 2022. [Online]. Available: \url{https://arxiv.org/abs/2202.03432}
\BIBentrySTDinterwordspacing

\bibitem{sicap}
\BIBentryALTinterwordspacing
J.~Silva-Rodríguez, A.~Colomer, M.~A. Sales, R.~Molina, and V.~Naranjo, ``\BIBforeignlanguage{en}{Going deeper through the gleason scoring scale: An automatic end-to-end system for histology prostate grading and cribriform pattern detection},'' \emph{\BIBforeignlanguage{en}{Computer Methods and Programs in Biomedicine}}, vol. 195, p. 105637, Oct. 2020. [Online]. Available: \url{https://linkinghub.elsevier.com/retrieve/pii/S016926072031470X}
\BIBentrySTDinterwordspacing

\bibitem{prostateGradingHarvard}
E.~Arvaniti, K.~S. Fricker, M.~Moret, N.~Rupp, T.~Hermanns, C.~Fankhauser, N.~Wey, P.~J. Wild, J.~H. Rueschoff, and M.~Claassen, ``Automated gleason grading of prostate cancer tissue microarrays via deep learning,'' \emph{Scientific reports}, vol.~8, no.~1, p. 12054, 2018.

\bibitem{BRACS}
N.~Brancati, A.~M. Anniciello, P.~Pati, D.~Riccio, G.~Scognamiglio, G.~Jaume, G.~De~Pietro, M.~Di~Bonito, A.~Foncubierta, G.~Botti \emph{et~al.}, ``Bracs: A dataset for breast carcinoma subtyping in h\&e histology images,'' \emph{Database}, vol. 2022, p. baac093, 2022.

\bibitem{bcnb}
F.~Xu, C.~Zhu, W.~Tang, Y.~Wang, Y.~Zhang, J.~Li, H.~Jiang, Z.~Shi, J.~Liu, and M.~Jin, ``Predicting axillary lymph node metastasis in early breast cancer using deep learning on primary tumor biopsy slides,'' \emph{Frontiers in oncology}, vol.~11, p. 759007, 2021.

\bibitem{spanhol2015dataset}
F.~A. Spanhol, L.~S. Oliveira, C.~Petitjean, and L.~Heutte, ``A dataset for breast cancer histopathological image classification,'' \emph{Ieee transactions on biomedical engineering}, vol.~63, no.~7, pp. 1455--1462, 2015.

\bibitem{msivsmss}
\BIBentryALTinterwordspacing
J.~N. Kather, ``Histological images for msi vs. mss classification in gastrointestinal cancer, ffpe samples,'' Feb. 2019. [Online]. Available: \url{https://zenodo.org/records/2530835}
\BIBentrySTDinterwordspacing

\bibitem{skinCancer}
K.~Kriegsmann, F.~Lobers, C.~Zgorzelski, J.~Kriegsmann, C.~Janssen, R.~R. Meli{\ss}, T.~Muley, U.~Sack, G.~Steinbuss, and M.~Kriegsmann, ``Deep learning for the detection of anatomical tissue structures and neoplasms of the skin on scanned histopathological tissue sections,'' \emph{Frontiers in Oncology}, vol.~12, p. 1022967, 2022.

\bibitem{tcgaUniform}
D.~Komura and S.~Ishikawa, ``Histology images from uniform tumor regions in tcga whole slide images,'' \emph{(No Title)}, 2020.

\bibitem{mpn}
U.~K.~M. Yusof, S.~Mashohor, M.~Hanafi, S.~M. Noor, and N.~Zainal, ``Histopathology imagery dataset of ph-negative myeloproliferative neoplasm,'' \emph{Data in brief}, vol.~50, p. 109484, 2023.

\bibitem{Kather2018_jx}
J.~N. Kather, N.~Halama, and A.~Marx, ``100,000 histological images of human colorectal cancer and healthy tissue,'' 2018.

\bibitem{Kather2016_fm}
J.~N. Kather, F.~G. Z{\"o}llner, F.~Bianconi, S.~M. Melchers, L.~R. Schad, T.~Gaiser, A.~Marx, and C.-A. Weis, ``Collection of textures in colorectal cancer histology,'' 2016.

\bibitem{sarcoma1}
H.~B. Arunachalam, R.~Mishra, O.~Daescu, K.~Cederberg, D.~Rakheja, A.~Sengupta, D.~Leonard, R.~Hallac, and P.~Leavey, ``Viable and necrotic tumor assessment from whole slide images of osteosarcoma using machine-learning and deep-learning models,'' \emph{PloS one}, vol.~14, no.~4, p. e0210706, 2019.

\bibitem{sarcoma2}
\BIBentryALTinterwordspacing
P.~Leavey, A.~Sengupta, D.~Rakheja, O.~Daescu, H.~B. Arunachalam, and R.~Mishra, ``Osteosarcoma ut southwestern/ut dallas for viable and necrotic tumor assessment,'' 2019. [Online]. Available: \url{https://www.cancerimagingarchive.net/collection/osteosarcoma-tumor-assessment/}
\BIBentrySTDinterwordspacing

\bibitem{renalCell1}
\BIBentryALTinterwordspacing
O.~Brummer, P.~Pölönen, S.~Mustjoki, and O.~Brück, ``\BIBforeignlanguage{en}{Integrative analysis of histological textures and lymphocyte infiltration in renal cell carcinoma using deep learning},'' \emph{\BIBforeignlanguage{en}{bioRxiv}}, p. 2022.08.15.503955, Aug. 2022. [Online]. Available: \url{https://www.biorxiv.org/content/10.1101/2022.08.15.503955v1}
\BIBentrySTDinterwordspacing

\bibitem{crcTp}
\BIBentryALTinterwordspacing
S.~Javed, A.~Mahmood, M.~M. Fraz, N.~A. Koohbanani, K.~Benes, Y.-W. Tsang, K.~Hewitt, D.~Epstein, D.~Snead, and N.~Rajpoot, ``Cellular community detection for tissue phenotyping in colorectal cancer histology images,'' \emph{Medical Image Analysis}, vol.~63, p. 101696, Jul. 2020. [Online]. Available: \url{https://www.sciencedirect.com/science/article/pii/S136184152030061X}
\BIBentrySTDinterwordspacing

\bibitem{chaoyang}
C.~Zhu, W.~Chen, T.~Peng, Y.~Wang, and M.~Jin, ``Hard sample aware noise robust learning for histopathology image classification,'' \emph{IEEE transactions on medical imaging}, vol.~41, no.~4, pp. 881--894, 2021.

\bibitem{tcgaTIL1}
J.~Saltz, R.~Gupta, L.~Hou, T.~Kurc, P.~Singh, V.~Nguyen, D.~Samaras, K.~R. Shroyer, T.~Zhao, R.~Batiste, J.~Van~Arnam, {Network, The Cancer Genome Atlas Research}, I.~Shmulevich, A.~U.~K. Rao, A.~J. Lazar, A.~Sharma, and V.~Thorsson, ``Tumor-infiltrating lymphocytes maps from {TCGA} {H\&E} whole slide pathology images,'' 2018.

\bibitem{tcgaTIL2}
J.~Saltz, R.~Gupta, L.~Hou, T.~Kurc, P.~Singh, V.~Nguyen, D.~Samaras, K.~R. Shroyer, T.~Zhao, R.~Batiste \emph{et~al.}, ``Spatial organization and molecular correlation of tumor-infiltrating lymphocytes using deep learning on pathology images,'' \emph{Cell reports}, vol.~23, no.~1, pp. 181--193, 2018.

\bibitem{tcgaTIL3}
K.~Clark, B.~Vendt, K.~Smith, J.~Freymann, J.~Kirby, P.~Koppel, S.~Moore, S.~Phillips, D.~Maffitt, M.~Pringle \emph{et~al.}, ``The cancer imaging archive (tcia): maintaining and operating a public information repository,'' \emph{Journal of digital imaging}, vol.~26, pp. 1045--1057, 2013.

\bibitem{snapFrozen}
J.~N. Kather, ``Histological images for {MSI} vs. {MSS} classification in gastrointestinal cancer, snap-frozen samples,'' 2019.

\bibitem{chatGpt}
OpenAI, ``Chatgpt,'' \url{https://chat.openai.com}, 2023, large language model. Accessed: 2025-02-18.

\bibitem{stainEffect}
S.~Lin, H.~Zhou, M.~Watson, R.~Govindan, R.~J. Cote, and C.~Yang, ``Impact of stain variation and color normalization for prognostic predictions in pathology,'' \emph{Scientific Reports}, vol.~15, no.~1, p. 2369, 2025.

\bibitem{stain1}
A.~Murmu and P.~Kumar, ``Automated breast nuclei feature extraction for segmentation in histopathology images using deep-cnn-based gaussian mixture model and color optimization technique,'' \emph{Multimedia Tools and Applications}, pp. 1--27, 2025.

\bibitem{stain2}
S.~Lin, H.~Zhou, M.~Watson, R.~Govindan, R.~J. Cote, and C.~Yang, ``Impact of stain variation and color normalization for prognostic predictions in pathology,'' \emph{Scientific Reports}, vol.~15, no.~1, p. 2369, 2025.

\bibitem{macenko}
M.~Macenko, M.~Niethammer, J.~S. Marron, D.~Borland, J.~T. Woosley, X.~Guan, C.~Schmitt, and N.~E. Thomas, ``A method for normalizing histology slides for quantitative analysis,'' in \emph{2009 IEEE international symposium on biomedical imaging: from nano to macro}.\hskip 1em plus 0.5em minus 0.4em\relax IEEE, 2009, pp. 1107--1110.

\end{thebibliography}

\appendix
\section{Caption Examples}

\begin{table}[t!]\centering
    \begin{adjustbox}{width=1\textwidth}
    \begin{tabular}{l|c}
    \toprule
     \bfseries Class & Synonym \\
    \bottomrule
      \multirow{4}{*}{normal} & "normal breast"\\
            & "standard breast"\\
            & "typical breast"\\
            & "non-cancerous breast" \\ 
    \midrule
      \multirow{4}{*}{pathologically benign} & "non-cancerous breast"\\
            & "normal cellular architecture"\\
            & "stable growth"\\
            & "physiologically typical breast"\\ 
    \midrule    
      \multirow{4}{*}{usual ductal hyperplasia} & "increased ductal cell growth"\\
            & "proliferation of ductal cells"\\
            & "non-cancerous ductal changes"\\
            & "common benign breast condition"\\ 
    \midrule
      \multirow{4}{*}{flat apithial atypia} & "ductal cell irregularities"\\
            & "mild epithelial atypia"\\
            & "benign cell changes"\\
            & "altered ductal architecture"\\ 
    \midrule
      \multirow{4}{*}{atypical ductal hyperplasia} &"abnormal ductal cell growth" \\
            & "epithelial cell irregularities"\\
            & "increased ductal atypia"\\
            &  "pre-cancerous breast condition" \\ 
    \midrule
      \multirow{4}{*}{ductal carcinoma in situ} & "localized ductal carcinoma"\\
            & "non-invasive breast cancer"\\
            & "pre-invasive ductal tumor"\\
            &  "atypical ductal proliferation"\\ 
    \midrule
      \multirow{4}{*}{invasive carcinoma} & "tumor invasion intos"\\
            & "malignant cell spread"\\
            & "infiltrative cancer growth"\\
            &  "aggressive tumor behavior"\\ 

    \end{tabular}
    \end{adjustbox}
    \caption{An example of the different class names used during caption ensembling. Each classname is substituted by one of the four corresponding synonyms and then integrated into the different caption templates.}
    \label{tab:captionEnsemble}
    \end{table}
 
\end{document}